\documentclass[final,1p,times]{elsarticle} 
\usepackage{graphicx} 
\usepackage{amssymb} 
\usepackage{amsthm} 
\usepackage{lineno} 

\journal{Nuclear Physics A} 
\begin{document} 

\begin{frontmatter} 


\title{Using Two- and Three-Particle Correlations to Probe Strongly Interacting Partonic Matter}

\author{Wolf G. Holzmann$^{a}$ for the PHENIX collaboration}

\address[a]{Columbia University and Nevis Laboratories, 
New York, NY 10027, USA}

\begin{abstract} 
The latest two- and three-particle correlation measurements obtained by the PHENIX
collaboration are presented. Three-particle correlations are consistent with the presence
of conical emission patterns in the data. Two-particle correlations relative to the collision
geometry reveal strong shape and yield modifications of the away-side jet, that depend on the orientation of the trigger particle with respect to the event plane.
A difference in the geometry dependence of the per trigger yields in the regions around $\Delta\phi \approx \pi$ and $\Delta\phi \approx \pi \pm 1.1$ can be understood by a different geometry 
dependence of jet energy loss and the medium response to the deposited energy.
\end{abstract} 

\end{frontmatter} 



\section{Introduction}
One of the most exciting results to have emerged from the Relativistic Heavy
Ion Collider (RHIC), is the large energy loss of high transverse momentum ($p_T$)
partons in the strongly interacting partonic matter created in heavy ion collisions
 \cite{raa}.  The fraction of energy transferred from the parton to the medium 
 is expected to depend on the gluon densities and the spacio-temporal extent of
 the interaction \cite{quench}. Therefore, jet energy loss is considered a promising
 tool to obtain a tomographic image of the partonic medium.
 Equally important is the medium excitation resulting from the energy transfer, since
 it may carry information on the transport properties of bulk QCD matter \cite{medmod}. 
 
 Azimuthal angular correlation measurements at intermediate $p_T$ reveal two 
 off-center away-side peaks at $\Delta\phi \approx \pi \pm 1.1$ \cite{ppg032}. These structures are
 generally attributed to the medium response to jets. Several possible mechanistic
 scenarios are being actively debated. At the center of the discussion is the creation
 of jet induced Mach shocks \cite{medmod}. Mach cones can provide information on the speed of sound
 ($c_S$) of strongly interacting partonic matter and are expected to be sensitive to the
 ratio of shear viscosity to entropy density ($\frac{\eta}{s}$). An important prerequisite
 to extracting medium properties from the data, however, is a solid understanding of the 
mechanistic details involved. Despite much recent progress, such an understanding
has not yet been achieved. Two important outstanding questions are wether conical
emission patterns do indeed exist in the data and how the underlying flow field may
influence the observed correlation structures.  

\section{Three-particle correlations}
\begin{figure}[ht]
\centering
%
%
\includegraphics[width=0.45\textwidth]{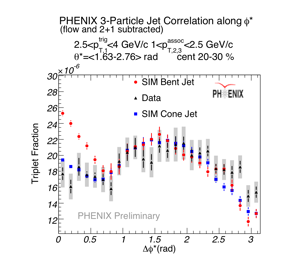}
\includegraphics[width=0.45\textwidth]{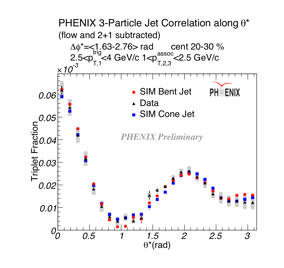}
%
\caption[]{Three-particle correlation functions for two charged hadrons at $1<p_{T,assoc}<2.5$ GeV/c
and a trigger hadron at $2.5<p_{T,trig}<4$ GeV/c in 20-30\% Au+Au collisions at $\sqrt{s_{NN}}=200$ GeV. Left: Azimuthal projection of two dimensional correlation function. Right: Radial projection of
two dimensional correlation function. }
\label{Fi:1}
\end{figure}

Three-particle correlation analyses provide added topological insights over and above 
the information content carried by two particle correlations. The PHENIX collaboration
has measured the correlation of two charged hadrons at $1<p_{T,assoc}<2.5$ GeV/c
associated with a trigger hadron at $2.5<p_{T,trig}<4$ GeV/c in Au+Au collisions
at $\sqrt{s_{NN}}=200$ GeV. To this end, a coordinate
system is defined for which the trigger particle orientation denotes the z-axis \cite{ajit}.
To the degree that the trigger particle approximates the jet axis, this coordinate system
is somewhat more natural to the investigated problem than the laboratory frame.
Figure \ref{Fi:1} shows the resulting azimuthal and radial projections of the
two dimensional correlation function for a centrality
selection corresponding to 20-30\% of the geometric cross-section. The
flow modulated background and the contributions from two-particle correlations 
have been subtracted following the method outlined in \cite{ajit}. The data (filled triangles)
is compared to toy Monte Carlo simulations of cone jets (filled squares) and bent
jets (filled circles). For the latter case, the away-side jets have been displaced in
azimuth to reproduce the signal in the two-particle correlation data. The comparison
indicates, that the observed correlation structures are consistent with expectations from 
conical emission patterns. An interplay of other contributions to the three-particle
correlation, however, cannot be ruled out. It is important to note, that the cone angle
inferred from the right hand panel of Fig. \ref{Fi:1} ($\theta_{cone} \approx 1$ rad) is comparable to the position of the
off-center peaks in the two-particle azimuthal correlation analyses.

The depicted three-particle correlation analysis integrates over the entire collision geometry. 
In order to  study the path-length dependence of jet-medium interactions, an analysis relative to
the event-plane was carried out.

\section{Two-particle correlations relative to the event-plane}
Jet energy loss depends on the initial gluon densities and the distance travelled
by the partons in the medium \cite{quench}. Similarly, the medium response to the energy transfer 
is expected to depend on the amount of matter that can be excited and the direction of the
parton with respect to the flow field.
Therefore, angular correlations relative to the reaction geometry
hold much promise to disentangle energy loss from medium excitation effects. They
can also serve to tightly constrain medium modification models.
To this end, we report relative azimuthal angle correlations of charged hadrons in the range 
1.0$<p_T<$2.0 GeV/c associated with a trigger hadron at 2.0$<p_T<$3.0 GeV/c, that has been oriented with respect to the 
event plane. The shape and yield of jet-induced correlations for both the near- and away-side jets are
then studied as a function of the relative azimuthal angle between the trigger particle and the event plane
($\phi_{S} = \phi_{trig}-\Psi$). 
The analysis follows the techniques outlined in Refs. \cite{ppg032,rplmath}. 
A newly installed reaction plane detector allows for the reconstruction of the event plane with 
excellent resolution ($\langle cos(2\Delta\Psi) \rangle \approx 0.74$ in mid-central collisions). This permits for differential studies relative to the collision geometry.

\begin{figure}[htp!]
\includegraphics[width=0.5\textwidth]{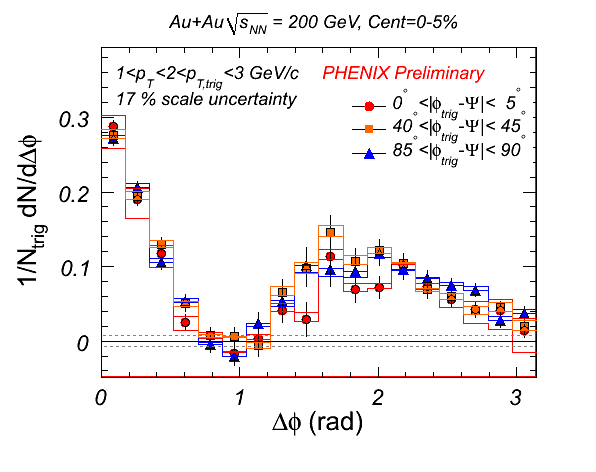}
\includegraphics[width=0.5\textwidth]{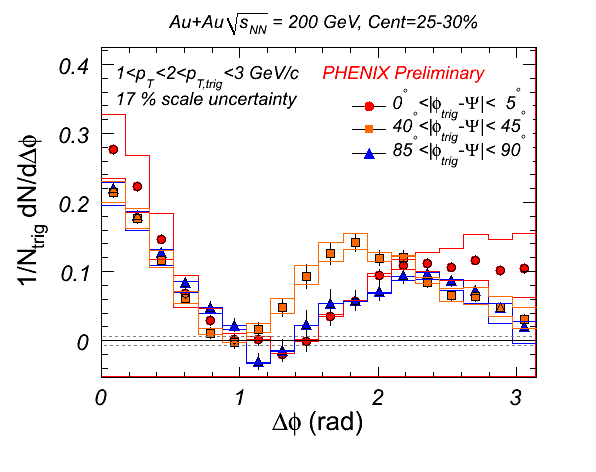}
\caption{Jet pair distributions for $1<p_{T,assoc}<2<p_{T,trig}<3$ GeV/c and 0-5\% and 25-30\% of the geometric cross section (top and bottom, repsectively). Shown are azimuthal
distributions where the trigger particle is oriented in-plane [filled circles, $|\phi_{S}|<5^{\circ}$] and out-of-plane [filled triangles, $85^{\circ}<|\phi_{S}|<90^{\circ}$], as well as a selection in between these two extremes [filled squares, $40^{\circ}<|\phi_{S}|<45^{\circ}$]. }
\label{Fi:2}
\end{figure}

Representative jet pair distribution per trigger particle are shown in 
Fig.~\ref{Fi:2} for 0-5\% and 25-30\% of the geometric cross section (left and right, repsectively). 
The cases where the trigger particle is oriented in-plane [filled circles, $|\phi_{S}|<5^{\circ}$] and out-of-plane [filled triangles, $85^{\circ}<|\phi_{S}|<90^{\circ}$], are contrasted with a selection in between these two extremes [filled squares, $40^{\circ}<|\phi_{S}|<45^{\circ}$]. 
The distributions have been folded into $|\Delta\phi| = 0-\pi$ and $|\phi_S| = 0-\pi/2$.

For the central events (Fig. \ref{Fi:2} left), the three jet pair distributions are essentially identical
within the stated uncertainties. This is to be expected, since central events do not result in strong 
reaction plane dependent path length changes. All three distributions show a broad away-side
peak that exhibits a minimum at $\Delta\phi = \pi$ and a maximum at $\Delta\phi \approx \pi-1.1$,
in accordance with earlier inclusive correlation measurements \cite{ppg032}.

For the mid-central selection (Fig. \ref{Fi:2} right), the away-side peak for the in-plane trigger
orientation and the out-of-plane trigger orientation both appear to have similar width. However,
the head region at $\Delta\phi \approx \pi$ seems to be suppressed in the out-of-plane
direction, while the distribution for
the in-plane trigger particle still shows a sizable yield at $\Delta\phi = \pi.$ This observation is
in qualitative agreement with expectations from jet suppression. For a di-jet system that is
aligned with the short axis of the overlap region (in-plane), very little jet quenching is expected.
By contrast, a jet opposite a trigger particle pointing perpendicular to the event plane has a
larger path length through the medium and is thus more likely to be quenched.

It is interesting to note, that the jet pair distribution for a trigger particle oriented at 
$40^{\circ}<|\phi_{S}|<45^{\circ}$ shows a much broader away-side than its in-plane and out-of-plane
counterparts. The yield is depleted
at $\Delta\phi=\pi$, but now shows an enhancement away from $\Delta\phi = \pi$ which peaks at $\Delta\phi \approx \pi - 1.3$. At present, it is unclear whether this merely reflects a geometry dependent shift
in the away-side peaks or perhaps an additional contribution at $\Delta\phi \approx \pi/2$. 

\begin{figure}[htp!]
\includegraphics[scale=0.4]{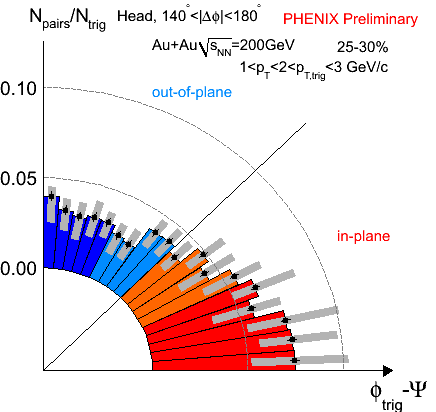}
\includegraphics[scale=0.4]{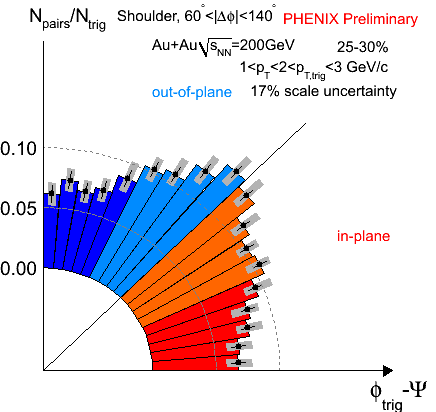}
\caption{Per-trigger-yield integrated over the azimuthal pair separations $140^{\circ}<|\Delta\phi|<180^{\circ}$ (left) and $60^{\circ}<|\Delta\phi|<140^{\circ}$ (right) plotted as a function of trigger particle
orientation relative to the event plane ($\phi_{trig}-\Psi$).}
\label{Fi:3}
\end{figure}

Figure \ref{Fi:2} suggests that two regions can be identified on the away-side. The region directly
opposite the trigger particle spanning $140^{\circ}<|\Delta\phi|<180^{\circ}$, which we will call
the "head" region and a "shoulder" region identified by $60^{\circ}<|\Delta\phi|<140^{\circ}$.
It is instructive to investigate the per-trigger-yield integrated in these regions as a function of the
trigger particle orientation with respect to the event plane. The result is given in Fig. \ref{Fi:3},
with the head yields depicted on the left and the shoulder yields on the right.
The head yields show a gradual depletion from in-plane to out-of-plane trigger particle orientations.
This is in agreement with qualitative expectations from jet-quenching models, since the path-length
opposite the trigger particle increases in the same direction. By contrast, the shoulder yields increase
from in-plane to reach a maximum at about $|\phi_{trig}-\Psi| \approx 45^{\circ}$. They fall again to
reach a value for the out-of-plane yield that is comparable to the in-plane case. The
correlation structures in the head region have predominantly been attributed to jet energy loss and the correlation
signals in the shoulder region are generally interpreted as arising mostly from the medium's response to this energy loss.
In such a picture, the different yield dependence can be understood by a difference in geometry dependence of jet energy loss and the medium response to the deposited energy.




\section*{Acknowledgments}
This work was supported by U.S. Department of Energy grant
DE-FG02-86ER40281.

\end{document}